\documentclass{tMOP2e}

\usepackage{epstopdf}
\usepackage{subfigure}
\usepackage{xcolor}

\theoremstyle{plain}

\theoremstyle{definition}

\theoremstyle{remark}

\begin{document}



\title{Multi-channel entanglement distribution using spatial multiplexing from four-wave mixing in atomic vapor}

\author{
\name{Prasoon Gupta\textsuperscript{a},
Travis Horrom\textsuperscript{a}$^{\ast}$\thanks{$^\ast$Corresponding author. Email: tshorrom@gmail.com},
Brian E. Anderson\textsuperscript{a},
Ryan Glasser\textsuperscript{b},
and Paul D. Lett\textsuperscript{a}}
\affil{\textsuperscript{a}National Institute of Standards and Technology
and Joint Quantum Institute, NIST and the University of Maryland,
Gaithersburg, Maryland 20899 USA;
\textsuperscript{b}Department of Physics and Engineering Physics, Tulane University, 6400 Freret Street, New Orleans, Louisiana 70118 USA}
\received{\today}
}

\maketitle

\begin{abstract}
Four-wave mixing in atomic vapor allows for the generation of multi-spatial-mode states of light containing many pairs of two-mode entangled vacuum beams.  This in principle can be used to send independent secure keys to multiple parties simultaneously using a single light source.  In our experiment, we demonstrate this spatial multiplexing of information by selecting three independent pairs of entangled modes and performing continuous-variable measurements to verify the correlations between entangled partners. In this way, we generate three independent pairs of correlated random bit streams that could be used as secure keys.  We then demonstrate a classical four-party secret sharing scheme as an example for how this spatially multiplexed source could be used.

\end{abstract}

\begin{keywords}
spatial multiplexing; four-wave mixing; secret sharing; entanglement
\end{keywords}

\section{Introduction}
Quantum key distribution (QKD) allows for the secure transfer of information by making use of quantum states which cannot be copied without detection \cite{Lo99, Grosshans03}. 
There have been numerous demonstrations of QKD in the continuous variable (CV) regime, both using squeezed or entangled states as well as modulated coherent states of light~\cite{Gottesman01, Cerf01, Xiaolong09, Grosshans02, Grosshans03}. While typical QKD implementations allow a link between two single points, a requirement for building a quantum network is to distribute the keys to several parties or share them among multiple users~\cite{QuCryptoReview, Phoenix95}.  

Secret sharing is a key distribution protocol that allows a sender to transmit information to several receivers in such a way that the receivers must work together to access the information~\cite{Shamir79}.   To protect against eavesdropping, multiple two-party QKD schemes can be used to securely transfer information to different receivers before a classical secret sharing protocol is carried out.  Methods also exist for sharing arbitrary quantum states among multiple parties at once~\cite{Hillery99, Cleve99, Tittle01, Chen05}  (called ``quantum secret sharing'' or ``quantum state sharing''), which allow for the quantum  cryptography and secret sharing to be combined in one secure step.   However these methods can be complicated and difficult to scale up to many users due to the requirements of multiparticle entanglement.  In this paper, we demonstrate quantum-assisted secret sharing as an example of using spatial multiplexing to establish independent secret keys with many users simultaneously through quantum channels.   Multiple pairs of entangled twin beams are used to generate and distribute pairs of correlated bits streams, which are then used in a classical secret sharing scheme.

The simplest way for a sender to establish multiple secure channels would be to carry out a separate QKD protocol with each receiver, either using multiple apparatuses, or a single apparatus used at different times. It was also shown that entanglement can be distributed to many parties for QKD using wavelength multiplexing in optical fibers~\cite{Lim08}, effectively increasing the number of available channels from a single resource.  In our experiment, we perform CV measurements on multi-spatial-mode light containing nonclassical correlations, demonstrating spatial multiplexing as a resource for distributing secure keys rather than wavelength multiplexing.  This allows for multiple keys to be established with multiple parties simultaneously, using a single source of light.  The two-mode squeezed vacuum state employed in our scheme does not require bright beams or modulation, as is often the case for CV QKD, but makes use of the intrinsic randomness of spontaneous emission to generate random keys.  While the present proof-of-principle experiment demonstrates the viability of spatial multiplexing as applied to secret sharing, proving the security of this scheme is beyond the scope of this work.  Security for two-mode squeezed vacuum states has previously been shown for two-party communication~\cite{Furrer12}, and there have been experimental demonstrations of CV QKD using single~\cite{Eberle13} and two-mode squeezing sources \cite{Madsen2012}.

A generic classical secret sharing protocol works as follows.  Alice wants to send a secret key to three agents, Bob, Charlie, and Diana, where all three recipients must work together to receive the key.  To accomplish this, Alice generates three pairs of correlated bit streams. Alice keeps one of each pair of bit streams, and the sum of these retained bit streams, modulo two, constitutes the key. A single bit stream is sent to each recipient, Bob, Charlie, and Diana, and each of these bit streams is statistically independent from the full key. Thus, none of the individual bit streams contains any information about the full key. Bob, Charlie, and Diana can meet and add their classical bit streams together to recover the key. 

To ensure that the parts of the key cannot be intercepted by an eavesdropper, Alice can send information via quantum communication channels and use quantum cryptography protocols with her agents~\cite{Silberhorn02-3dB, Grosshans03}.  In our experiment, the partial keys are generated by a nondeterministic quantum process and transmitted via nonclassical states of light.  Based on the measurements on the quantum states they receive, each party generates a  bit stream to be used in the classical secret sharing protocol described above.

\begin{figure*}
\begin{center}
\includegraphics[width=.95\textwidth]{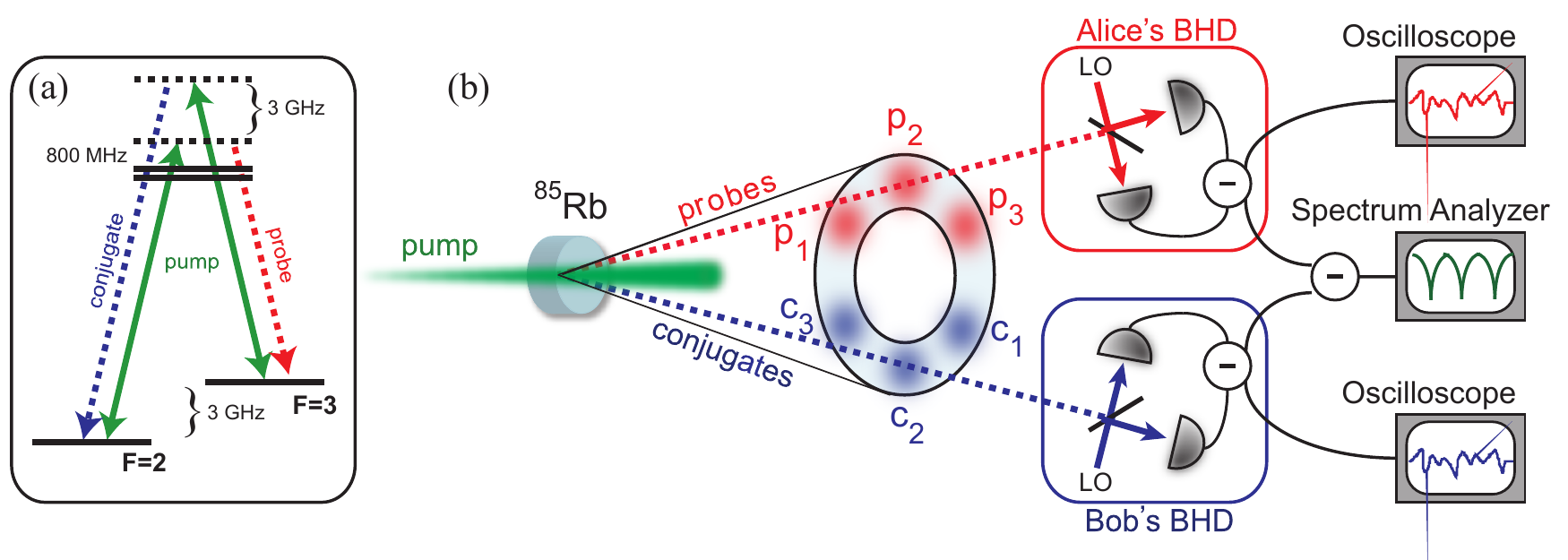}
\caption{(Color online) (a) Atomic level diagram for the D1 line of $^{85}$Rb. (b) Partial sketch of experimental setup. For clarity, we only show two spatial mode paths and homodyne detectors.  In the actual experiment, six individual spatial modes from the generated light state are simultaneously aligned to six balanced homodyne detectors (BHDs) for quadrature measurements. LO denotes the local oscillators.}
\label{fig:setup}
\end{center}
\end{figure*}

We use four-wave mixing (4WM) in hot $^{85}$Rb vapor to generate two-mode squeezed vacuum states of the electromagnetic field~\cite{Boyer08}.  The multi-spatial-mode nature of this field gives us multiple correlated pairs of probe and conjugate modes (vacuum twin beams).  Alice performs continuous variable quadrature measurements on three of the probe modes and bins the results to produce three random bit streams used to generate her key.  At the same time, she sends the three conjugate modes to Bob, Charlie, and Diana, who perform similar quadrature measurements. When the correlated quadratures are measured, the three agents can combine their resulting bit streams to recover a key which is correlated with Alice's key, as discussed below. By examining the joint quadratures of a probe-conjugate pair, the squeezing of the light states can be measured, verifying that nonclassical states are transmitted.

\section{Experiment}
The 4WM setup, similar to that described in~\cite{Boyer08}, is shown in  Fig.~\ref{fig:setup} for one set of quadrature measurements. The full apparratus consists of a Ti:sapphire laser pumping the 4WM nonlinear medium to generate the vacuum twin beams, and six balanced homodyne detectors (BHDs) to record the quadratures of three probe and three conjugate modes.  The 400 mW pump beam, with $1/e^2$ beam diameter of 700 $\mu$m, is detuned 800 MHz to the blue from the $^{85}$Rb D1 $F_g=2\rightarrow F_e=3~$ transition and passes through a vapor cell of $^{85}$Rb heated to 114$^\circ$~C.  To align the chosen vacuum modes into the BHDs, a portion of the pump light is double-passed through a 1.5~GHz acousto-optical modulator (AOM) giving us a seed probe beam shifted from the pump by 3 GHz, as shown in Fig.~\ref{fig:setup}a. This probe beam crosses the pump beam at a small angle in the vapor cell to seed the 4WM, resulting in bright twin beams.  This input seed is divided and aligned to three distinct azimuthal angles with respect to the pump to produce three spatially distinct pairs of twin beams.  To produce squeezed vacuum beams, the probe seeds are blocked.  A second seeded 4WM process is prepared in the same vapor cell to provide bright local oscillators (LO) for homodyne detection. The BHDs contain matched photodiodes with 95~$\%$ quantum efficiency, and we maintain interference visibilities of 96$~\%$ to 97~$\%$.  In the secret sharing arrangement, Alice controls the three BHDs measuring the probe beams, while Bob, Charlie, and Diana control the conjugate BHDs.

The output from each of the BHDs represents either a measurement of the amplitude ($\hat{X}_p$,  $\hat{X}_c$) or phase quadrature ($\hat{Y}_p$, $\hat{Y}_c$) for the probe or conjugate modes, as indicated by the subscript. The measured quadrature is selected by controlling the relative phase of the LO by changing the LO path length with piezo-mounted mirrors.  The outputs from the BHDs are digitized, saved, and used to generate bit streams to build a key. At the same time, signals from correlated probe and conjugate BHDs are subtracted and sent to spectrum analyzers.  This allows us to monitor the noise of the joint quadratures of the correlated beams, $(\Delta\hat{X}_-)^2= (\Delta (\hat{X}_p-\hat{X}_c)/\sqrt{2})^2$.  Noise locking~\cite{McKenzie05} is used to stabilize the phases of the LOs and observe the squeezed noise of the amplitude quadrature, $\Delta \hat{X}_-$, below the shot noise limit (SNL).  
In this experiment, we observed squeezing of ($-2.0\pm 0.2$, $-2.3\pm 0.1$, and $-2.1\pm 0.2$) dB as compared to the shot noise, at the sideband frequency of 1~MHz, for the three channels respectively. The uncertainties represent the standard deviation over ten noise power measurements.
We note that some QKD protocols require a minimum of -3~dB squeezing~\cite{Usenko2011}, and although not reached in this experiment, squeezing levels exceeding this limit have been previously reported for our 4WM system~\cite{Boyer08}.  

In order to extend the present experiment to demonstrate provable security, all six of the local oscillator phases would have to be chosen independently and randomly for each measurement~\cite{Madsen2012}. In a true QKD demonstration, no noise locking scheme would connect Alice to the receivers; an alternate method would need to be used to derive a phase reference for choosing the quadratures.  For example, phase-locked local oscillators could be distributed along a common path in an orthogonal polarization to the probe and conjugate beams for free space key distribution~\cite{Cerf01}.

For the creation of a key, we wish to use the continuous-variable quadrature measurements to generate random bit strings.  The probe and conjugate beams are Gaussian states with noise in excess of shot noise due to spontaneous emission in the 4WM cell.  Since spontaneous emission is a nondeterministic quantum process, the noise is random. The signal from each BHD is saved and filtered, as discussed below. After filtering, we are left with a distribution of the noise from each BHD signal symmetric about zero voltage, and we can assign a `1' if the integration within a specified time slice is positive, or a `0' if it is zero or negative.   This method is similar to that used by Gabriel \emph{et al.}~\cite{Gabriel10}, where they use the noise of a coherent vacuum state to generate random numbers. Here we use the noise of the probe and conjugate squeezed vacuum modes to generate random numbers.  Other methods of discretization exist which could be used to increase the bit rate and help differentiate between classical and quantum noise, but for simplicity we use only the positive-negative binning~\cite{Gabriel10}.   
				
\begin{figure}
\begin{center}
\includegraphics[width=.5\columnwidth]{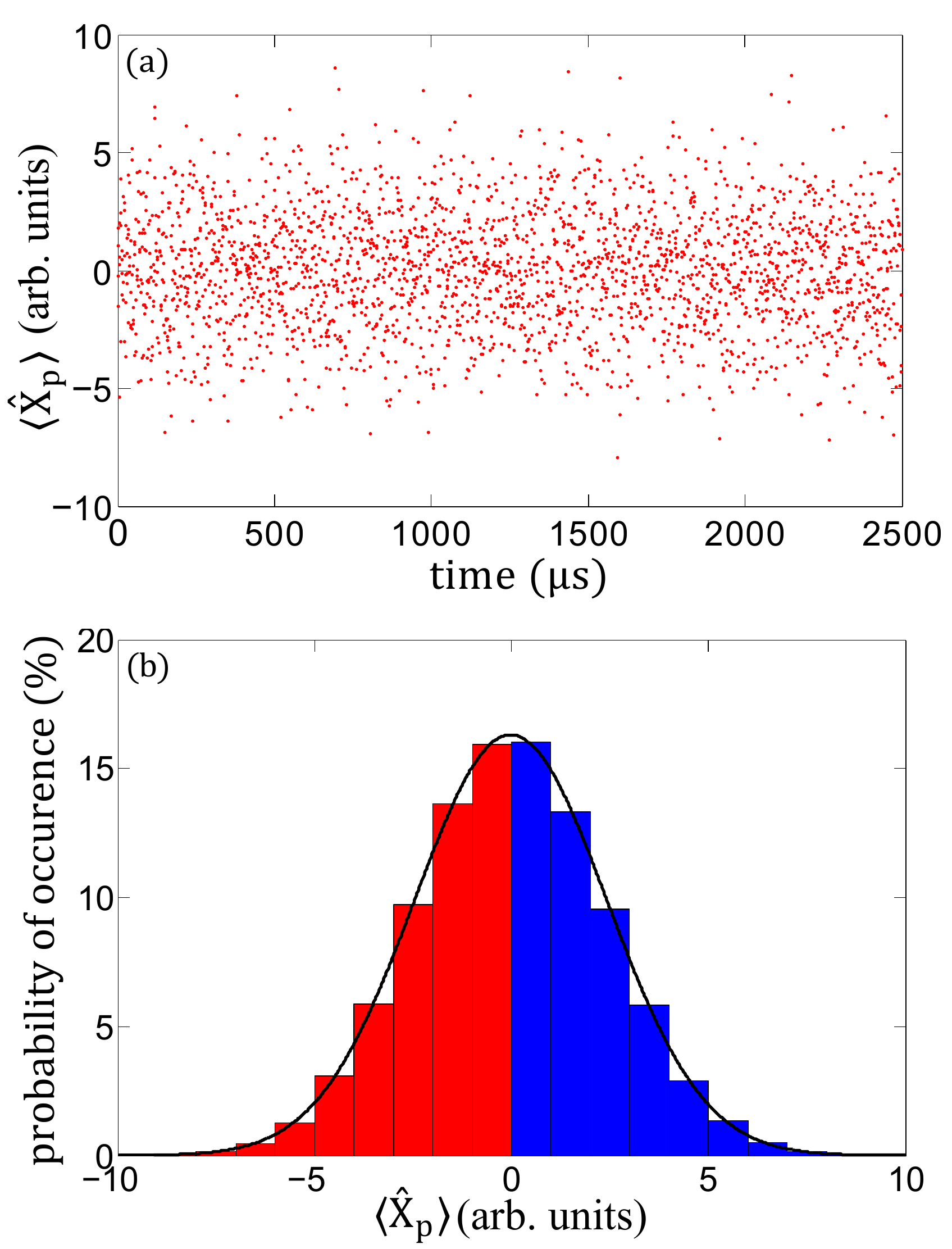}
\caption{(Color online) (a) Typical measurements of the X quadrature of a probe beam versus time with an integration time of 500~ns, buffer time of 500~ns, and bandpass filter of 15~kHz to 2~MHz.  (b) Binned probability distribution  of quadrature data shown with a Gaussian fit. Blue (dark) is `1' and red (light) is `0' .}
\label{fig:probe_quadrature}
\end{center}
\end{figure}

For secure communication to be possible, we must generate a key from signals that have non-classical correlations since any classical signal could be vulnerable to eavesdropping.  For this reason, we apply a sharp bandpass filter to the homodyne data from 15~kHz to 2~MHz.  This excludes low frequencies, which are dominated by technical noise, as well as higher frequencies where our squeezing begins to degrade.  We integrate over 500~ns time slices for each bit and exclude data from 500~ns buffer times between successive time slices to ensure that each bit is uncorrelated from the last, giving a bit generation rate of 0.9~Mbit/sec.  An example of the processed data for one probe mode detection is shown in Fig.~\ref{fig:probe_quadrature}a, with Fig.~\ref{fig:probe_quadrature}b highlighting the Gaussian distribution of noise.  We use bit streams of 90\,000 points for each probe and conjugate beam to generate the keys in this demonstration.
				
\section{Results}
To test the randomness of the generated bit streams, we use the National Institute of Standards and Technology Statistical Test Suite~\cite{NistTestSuite}, which includes multiple stringent randomness tests, 11 of which are applicable to our bit streams.  When averaged and filtered as described above, the six bit streams generated from the probe and conjugate quadrature measurements passed 64 out of 66 (11 tests for each bit stream) applicable tests. Passing a test means the sequence is random with a confidence of 99~\% with regard to the criteria of the particular test. We note that changing the integration time or the filtering parameters can affect the performance on the randomness tests. In the present experiment, these parameters were chosen simply to exclude technical noise at low and high frequencies. 
				
\begin{figure}
\begin{center}
\includegraphics[width=.5\columnwidth]{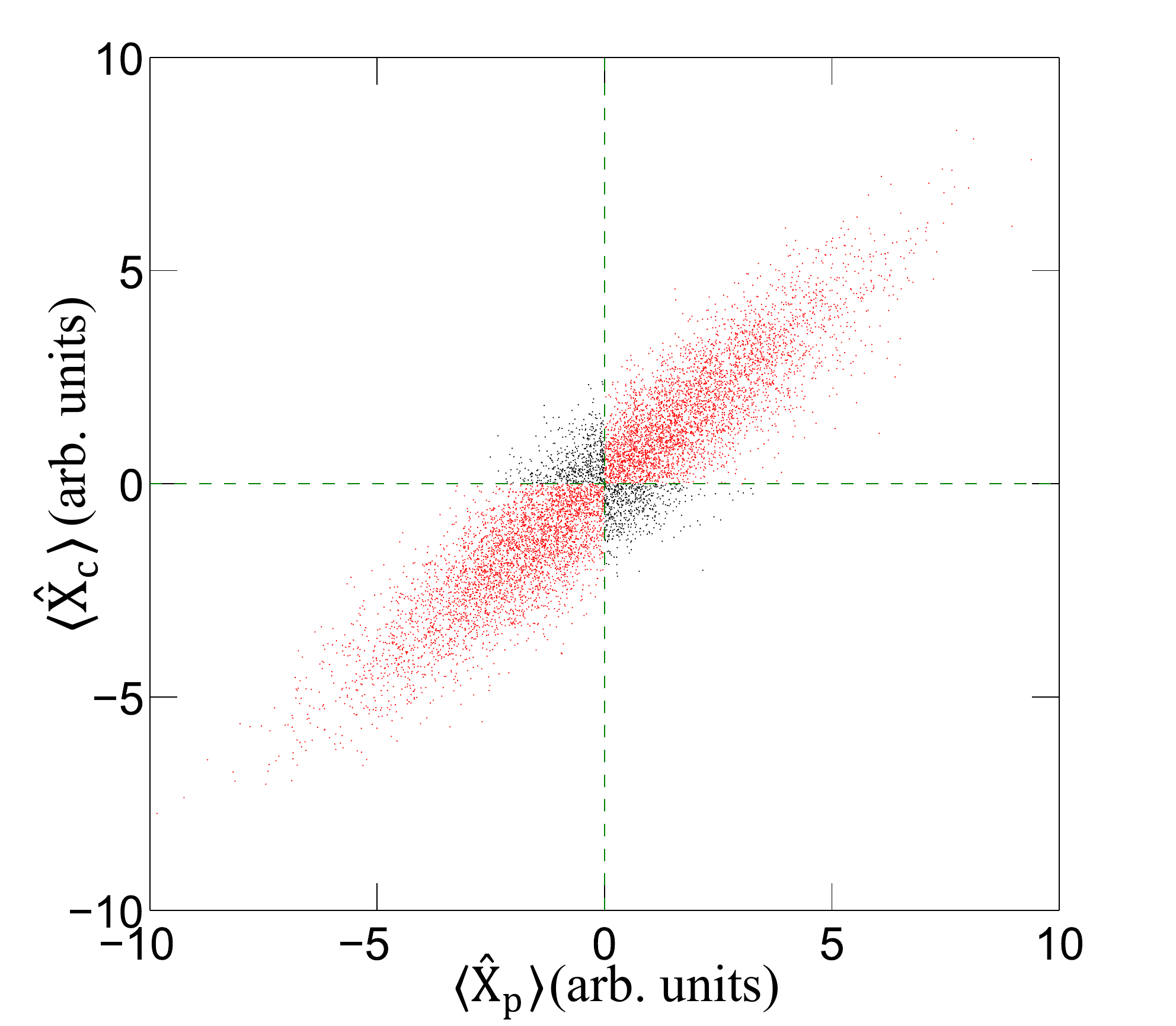}
\caption{(Color online) Experimental scatter plot of the average value of the probe quadrature $\hat{X}_p $ as a function of the corresponding conjugate quadrature $ \hat{X}_c $.  Data points in quadrants I and III (red, light) indicate correlated bits while those in II and IV (black, dark) will lead to error bits. Measured bit agreement rate for this path is $87~\%$. }
\label{fig:Wigner}
\end{center}
\end{figure}

The level of correlation between probe and conjugate modes determines the overall level of bit-wise agreement between discretized signals. Each conjugate beam, detected by Bob, Charlie, and Diana, should be correlated with its corresponding twin probe, detected by Alice, but uncorrelated from the other conjugate and probe modes.  In this case, the key created by adding together the probe bit streams will be completely uncorrelated with each individual conjugate bit stream, as well as any combination of bit streams not including the full set.  The streams are compared bit-wise and the agreement rates are shown in Table~\ref{Tab:success_rates}, where $100~\%$ indicates perfect agreement of the bit streams, and $50~\%$ indicates no correlation. In Fig. \ref{fig:Wigner} we show a scatter plot of the experimental amplitude quadrature of the probe $\hat{X}_p$ versus that of the conjugate $\hat{X}_c$. This distribution shows the squeezing ellipse and corresponds to an 87~\% correlation. We also calculate the theoretical Wigner function using the measured squeezing and excess noise levels of the probe and conjugate beams~\cite{Simon00, Laurat05}. From this we can find a theoretical bit agreement rate of 84~\% The slightly lower theoretical value is likely due to filtering the data and an underestimate of the excess noise of the twin beams used in calculating the theoretical value.		

\begin{table}
\tbl{Percent bit agreement between probe (P$_i$) and conjugate (C$_i$) bit streams, each containing 90\,000 points. The subscripts 1, 2, and 3 show the set of beams from which the bit streams were generated. Error bars are the standard deviation of 10 samples of 9\,000 points each.
}
{\begin{tabular}{ |c | c | c | c | }
  \hline
   & C$_{1}$ & C$_{2}$ & C$_{3}$  \\  \hline                 
  P$_{1}$ & $\mathbf{86.9\pm 0.3 }$  & $50.0\pm 0.4$ & $49.8\pm 0.4$ \\ \hline
  P$_{2}$ & $50.1\pm 0.5$ & $\mathbf{88.7\pm 0.4}$ & $49.8\pm 0.5$ \\ \hline
  P$_{3}$ & $49.8\pm 0.6$ & $49.9\pm 0.5$ & $\mathbf{88.6\pm 0.5}$ \\ \hline 
	\end{tabular}}
\label{Tab:success_rates}
\end{table}

When the three conjugate bit streams are added together, we achieve a total bit agreement rate of $72~\%$ for the reconstructed key as compared to Alice's key, formed by adding the three probe bit streams.  As expected, no subset of the full key reconstruction gives any information about the original key, as shown in Table \ref{Tab:total_key}.

\begin{table}
\tbl{Percent bit agreement between conjugate bit stream combinations and the full key: P$_1$+P$_2$+P$_3$.}
{\begin{tabular}{ |c | c |  }
  \hline
  & key$=$P$_{1}$ + P$_{2}$ + P$_{3}$  \\  \hline                 
  C$_{1}$                & $49.8\pm 0.5$ \\ \hline
  C$_{2}$                & $50.0\pm 0.4$ \\ \hline
  C$_{3}$                & $50.0\pm 0.4$ \\ \hline
	C$_{1}$+C$_2$         & $49.8\pm 0.6$ \\ \hline 
	C$_{1}$+C$_3$         & $50.0\pm 0.7$ \\ \hline
	C$_{2}$+C$_3$         & $50.3\pm 0.3$ \\ \hline
	C$_{1}$+C$_2$+C$_3$  & $\mathbf{72.0\pm 0.5}$ \\ \hline
	\end{tabular}}
\label{Tab:total_key}
\end{table}

\section{Summary} 
We have presented the first demonstration of spatial-mode multiplexing for entanglement distribution and have also demonstrated the suitability of 4WM in atomic vapors as a resource for QKD schemes. This light source allows for multiple sets of entangled modes to be distributed simultaneously, increasing the number of possible users or the information transfer rate compared with a single-mode system.  Secret sharing is one example of a multi-user cryptography protocol where 4WM could be applicable.  Improvements can be made by optimizing the 4WM source for higher levels of squeezing as well as by using error correction and privacy amplification~\cite{Eberle13}. The present experiment should be scalable to many more users, because it has been estimated that the output state of this 4WM process contains roughly 100 distinguishable spatial modes~\cite{Boyer08Imaging}.

\section*{Acknowledgements}
  The authors thank Kevin Jones and Quentin Glorieux for useful discussions.

\section*{Funding}	
This work was supported by the Air Force Office of
Scientific Research, grant number FA9550-1-0035.

\bibliographystyle{tMOP}
\bibliography{bibliography}

\end{document}